# Three dimensional collective charge excitations in electron-doped cuprate superconductors

M. Hepting[1], L. Chaix[1], E. W. Huang[1,2], R. Fumagalli[3], Y. Y. Peng[3,†], B. Moritz[1], K. Kummer[4], N. B. Brookes[4], W. C. Lee[5], M. Hashimoto[6], T. Sarkar[7], J. F. He[1,‡], C. R. Rotundu[1], Y. S. Lee[1], R. L. Greene[7], L. Braicovich[3,4], G. Ghiringhelli[3,8], Z. X. Shen[1*], T. P. Devereaux[1*], and W. S. Lee[1*]

[1]Stanford Institute for Materials and Energy Sciences, SLAC National Accelerator Laboratory and Stanford University, 2575 Sand Hill Road, Menlo Park, California 94025, USA

[2]Department of Physics, Stanford University, Stanford, California 94305, USA

[3]Dipartimento di Fisica, Politecnico di Milano, Piazza Leonardo da Vinci 32, I-20133 Milano, Italy

[4]European Synchrotron Radiation Facility (ESRF), BP 220, F-38043 Grenoble Cedex, France

[5]Department of Physics, Binghamton University, Binghamton, New York 13902, USA

[6]Stanford Synchrotron Radiation Lightsource, SLAC National Accelerator Laboratory, Menlo Park, CA 94025, USA

[7]Center for Nanophysics and Advanced Materials, Department of Physics, University of Maryland, College Park, Maryland 20742, USA

[8]CNR-SPIN, Politecnico di Milano, Piazza Leonardo da Vinci 32, I-20133 Milano, Italy

Correspondence to: leews@stanford.edu, zxshen@stanford.edu, tpd@stanford.edu

Present addresses:
†Department of Physics and Seitz Materials Research Lab, University of Illinois, Urbana, IL 61801, USA
‡Department of Physics, University of Science and Technology of China, Hefei, Anhui 230026, China

10/28/18

**High temperature cuprate superconductors consist of stacked $CuO_2$ planes, with primarily two dimensional electronic band structures and magnetic excitations [1,2], while superconducting coherence is three dimensional. This dichotomy highlights the importance of out-of-plane charge dynamics, believed to be incoherent in the normal state [3,4], yet lacking a comprehensive characterization in energy-momentum space. Here, we use resonant inelastic x-ray scattering (RIXS) with polarization analysis to uncover the pure charge character of a recently discovered collective mode in electron-doped cuprates [5-7]. This mode disperses along both the in- and, importantly, out-of-plane directions, revealing its three dimensional nature. The periodicity of the out-of-plane dispersion corresponds to the $CuO_2$ plane distance rather than the crystallographic *c*-axis lattice constant, suggesting that the interplane Coulomb interaction drives the coherent out-of-plane charge dynamics. The observed properties are hallmarks of the long-sought acoustic plasmon, predicted for layered systems [8-13] and argued to play a substantial role in mediating high temperature superconductivity [13-15].**

The charge dynamics of systems with periodically stacked quasi-two dimensional (2D) conducting planes are drastically affected in the presence of poorly screened interplane Coulomb interactions. In a simple layered electron gas model with conducting planes separated by dielectric spacers [8-10], the dispersion of plasmons, the collective electronic modes of the charge dynamics, evolves from optical-like to acoustic-like as a function of out-of-plane momentum $q_z$ [Fig. 1(a)], a behavior distinct from that in either pure 2D or isotropic 3D systems. For superconducting cuprates, similar charge dynamics have been postulated since they consist of conducting $CuO_2$ planes stacked along the *c*-axis with poor out-of-plane screening [11-13]. While plasmons were observed in various spectroscopic studies at the Brillouin zone center [4,16,17] and by transmission electron energy loss spectroscopy (EELS) typically exploring in-plane energy-momentum dispersions at $q_z = 0$ [18], there is no information on the $q_z$ dependence. Experimental evidence of this previously undetected component and its characterization in energy-momentum space can shed new light on long-standing proposals that connect out-of-plane charge dynamics to superconductivity. For instance, it has been proposed that 20% of the observed value of the cuprates' high $T_c$ can be attributed to the presence of acoustic plasmons [11-13] and that the energy stored in interplane Coulomb interactions is related to the energy savings associated with the superconducting transition [14,15].

In this letter, we focus our attention on the enigmatic "zone center" excitation previously discovered by RIXS in the electron-doped cuprate $Nd_{2-x}Ce_xCuO_4$ (NCCO) ($x = 0.15$) [5,6] and $Sr_{1-x}La_xCuO_4$ [7]. In a different family of electron-doped cuprates $La_{2-x}Ce_xCuO_4$ (LCCO) (x = 0.175), we resolve spectral features closely similar to NCCO at representative in-plane momentum transfer positions $q//$ [Figs. 1(b), (c)], suggesting the universality of this collective mode in electron-doped cuprates. Speculation about its origin included intra-band transitions [5], collective modes of a quantum phase [6], and plasmons [19]. Although the mode has been suspected to be of charge character, the sensitivity of RIXS to both charge and magnetic excitations [20-22] obscured a definitive assignment in previous experiments.

We first identify the character of this excitation by determining the associated magnetic and charge contributions to the RIXS spectra. This can be uniquely achieved by resolving the polarization of both the incident and scattered photons [21]. Namely, magnetic excitations flip spins and necessarily change the angular momentum of the photons in the scattering process, *i.e.* contribute to the crossed-polarization channel (σπ or πσ). Conversely, charge excitations preserve the angular momentum of the photon and contribute to the parallel polarization channel (σσ or ππ). Figs. 1(d) and (e) show polarization-resolved RIXS spectra at for two different in-plane momenta. At q// = (0.045, 0) the features of the zone center excitation are fully suppressed in the crossed polarization (σπ) channel and the spectrum solely contains the parallel polarization (σσ) contribution. For larger momentum transfer q//= (0.095, 0) the mode disperses toward higher energy (~ 0.8 eV) [Fig. 1(e)] and the well-known paramagnon excitation emerges on a lower energy scale (~ 0.3 eV). As expected, the paramagnon yields spectral weight in both channels due to the mixture of single spin flip excitation, and double spin-flip and incoherent particle-hole charge excitations, with the latter becoming significant with increasing doping concentration [20,21,23]. Importantly, the zone center excitation, which is separated in energy from the paramagnon, still appears only in the parallel polarization channel. Thus, we conclude that the zone center excitation is a pure charge mode.

A second key insight can be obtained from a comprehensive mapping of the energy-momentum dispersion in all three dimensions of reciprocal space, in contrast to prior RIXS experiments that explored the projected in-plane momentum without focusing on the $q_z$ dependence [6-8]. Figs. 2(a) and (b) show the RIXS intensity maps as a function of momentum transfer along the *hh*- and *h*-directions (*i.e.* along (0, 0, *l*)-(*h*, *h*, *l*) and (0, 0, *l*)-(*h*, 0, *l*), respectively) at *l* = 1 and *l* = 1.65. We denote *h, k, l* in reciprocal units ($2\pi/a$, $2\pi/b$, $2\pi/c$), where *a, b* = 4.01 Å and *c* = 12.4 Å are the lattice constants of LCCO with *x* = 0.175. For both *l,* the zone center excitation exhibits an almost linear dispersion emanating away from the zone center (0, 0, *l*) along both the *hh*- and *h d*irections. Surprisingly, while the dispersion of the paramagnon remains unchanged for different *l* values (as expected for the quasi-2D magnetic structure of cuprates), the dispersion of the zone center excitation becomes steeper and increasingly separated from the paramagnon branch at larger *l*. This behavior is highlighted in Fig. 2(c). The raw data itself also clearly shows such an *l*-dependence: the spectral peak shifts to higher energy and becomes broader at the higher *l*-value for a given in-plane momentum [Figs. 2(d), (e)].

Evidently, such finding calls for a more detailed investigation along the out-of-plane direction. Fig. 3(a) displays the energy-momentum dispersion as a function of $l$ at fixed in-plane momentum transfer (0.025, 0) near the in-plane zone center. As anticipated, the dispersion is symmetric around the out-of-plane zone center $l = 1$ since it is a high symmetry point in reciprocal space. Remarkably, with further increasing $l$, the zone center excitation continues dispersing toward high energy, insensitive to the next high symmetry point at $l = 1.5$. In fact, we also observe the same peculiar behavior in our new momentum-resolved RIXS measurements on NCCO (see Supplementary Fig. 7), suggesting a universal origin to the three dimensionality of the charge mode in an otherwise layered quasi-2D material.

These results can be rationalized by doubling the out-of-plane Brillouin zone size, implying that the crystallographic unit cell with lattice constant $c$ does not set the periodicity of the zone center excitation, but rather $d = c/2$: the nearest neighbor $CuO_2$ plane spacing [Fig. 3(b)]. Hence, a new index $l^*$ in units of $2\pi/d$ appropriately describes the Brillouin zone "felt" by the zone center excitation and establishes the proper periodicity of the dispersion. An obvious candidate which can induce such a Brillouin zone "reconstruction" in a quasi-2D system is the interplanar Coulomb interaction.

These striking features are reminiscent of acoustic plasmon bands theoretically proposed in the aforementioned layered electron gas model [Fig. 1(a)], which provides a qualitative fit to the observed zone center excitation (See Supplementary Information and Supplementary Fig. 8 for the fits). To demonstrate the behavior of collective charge dynamics beyond this weakly correlated layered electron gas model, we perform determinant quantum Monte Carlo (DQMC) calculations for a 2D three-band Hubbard model with an electron doping of $x = 0.18$, and incorporate three dimensionality through interplane Coulomb interactions using a random phase approximation (RPA)-like formalism (see Methods). Therewith, we obtain the collective charge response in form of the loss function (-Im 1/ε) vs. out-of-plane momentum transfer $l^*$. As shown in Fig. 3(c), the calculated dispersions along the $l^*$-direction at in-plane momenta accessible in our 16 x 4 cluster exhibit the same qualitative behaviors observed in both LCCO [Fig. 3(a)] and NCCO (Supplementary Fig. 7). We note that recent calculations using different methods and models have also demonstrated $l^*$-dependent plasmon bands [19,24]. While RIXS is not simply proportional to (-Im 1/ε) due to the resonant process [21,25], it nevertheless can contain critical information on the loss function, such as the dispersion of excitations. Thus, the agreement between our data and theory lends strong support to attributing the zone center mode to a plasmon excitation, with momentum resolved RIXS providing unique access to its acoustic bands.

Interestingly, the data shown in Figs. 2 and 3(a) indicate that the plasmon peak broadens when approaching the equivalent zone center along the $l^*$-direction (see also Supplementary Figs. 5(c) and 7(c)). At $q = (0.025, 0, l^* = 0.925)$, i.e. close to the equivalent zone center $(0, 0, l^* = 1)$, the plasmon linewidth of approximately 0.5 eV [Fig. 3(a)] is similar to previous transmission EELS reports [18] and optical conductivity measurements at (0, 0, 0) [17]. The increasing incoherence of the mode near the zone-center is consistent with incoherent charge dynamics inferred from $c$-axis optical conductivity [3,4]. However, such linewidth evolution appears to be different from the Landau quasi-particle picture in which the plasmon peak should be sharpest at the zone center where the continuum is minimal and well separated from the plasmon. Other mechanisms, such as the presence of non-Fermi liquid behavior, polar interlayer electron-phonon coupling [26], and/or Umklapp scattering [27], may reconcile these observations.

We now turn to the investigation of the acoustic plasmon bands in LCCO as function of the Ce doping concentration, or carrier density $x$. As shown in Figs. 4(a) and (b) the plasmon bands exhibit detectable doping dependence. In a naive picture, the plasmon energy is expected to be zero in the absence of carriers and increases proportionally to $\sqrt{x/m^*}$, where $m^*$ is the effective electron mass. Consistent with such expectation the mode energies of $x$ = 0.11 to ~ 0.15 increases linearly with $\sqrt{x}$ [Fig. 4(c)], further substantiating the attribution of the zone center excitation to a plasmon. For higher dopings, the rate of increase slows down and appears to hit a plateau for $x$ = 0.17 and 0.18. This observation suggests a possible variation in the band dispersion or Fermi-surface approximately at $x$ ~ 0.15. Recent Hall-effect and magnetoresistivity measurements on LCCO have indicated Fermi surface reconstruction due to antiferromagnetic correlations ending at around $x$ ~ 0.14 [28,29], corroborating our observation.

Our results reveal comprehensive information about the charge dynamics of electron-doped cuprates. The collective charge modes propagate in all three dimensions of reciprocal space with different in- and out-of-plane dynamics. They are not quasi-2D, as one may have anticipated from the single-particle spectrum, and the out-of-plane momentum dependence highlights the importance of interplane Coulomb interaction. Significant structural and electronic analogies between electron- and hole-doped materials imply a similar three dimensionality would be expected in the latter's charge dynamics. While previous RIXS studies on hole-doped compounds primarily focused on magnetic, orbital, and other high energy excitations [22], few investigated the region of energy-momentum space necessary to identify and characterize the acoustic plasmon [30]. More detailed RIXS measurements on hole-doped compounds, both single and multi-layer systems with higher $T_c$, can provide clarity on the role of interplane Coulomb interaction and acoustic plasmons in achieving high temperature superconductivity [14,15]. Moreover, our demonstration of using RIXS to gain insights into plasmonics of layered systems in three dimensional reciprocal space, can open new routes for studying heterostructures of 2D layered materials and stacked graphene multilayers.

**Methods**

The *c*-axis oriented LCCO thin films with Ce concentrations of $x$ = 0.11, 0.13, 0.15, 0.17, 0.175, and 0.18 were fabricated on (100) SrTiO3 substrates by pulsed laser deposition utilizing a KrF excimer laser. The annealing process was optimized for each $x$. The superconducting $T_c$ of the $x$ = 0.11 and 0.13 films is ~ 30 K and ~ 22 K, respectively. The $x$ = 0.175 and 0.18 films did not show a superconducting transition. The NCCO single crystal with Ce concentration $x$ = 0.15 was grown by traveling-solvent floating-zone method in $O_2$ and annealed in Ar at 900 °C for 10 hours. The $T_c$ is ~ 26 K.

The RIXS measurements were performed at beamline ID32 of the European Synchrotron Radiation Facility (ESRF, Grenoble, France) using the high-resolution ERIXS spectrometer. The scattering angle $2\Theta$ can be changed in a continuous way from 50° to 150°. The samples were mounted on the 6-axis in-vacuum Huber diffractometer/manipulator and cooled to ~ 20 K. The RIXS data were obtained with incident σ polarization (perpendicular to the scattering plane, high through-put configuration). The incident photon energy was tuned to the maximum of the Cu $L_3$ absorption peak at ~ 931 eV. The energy resolution was approximately $\Delta E$ ~ 60 meV for the $x$ = 0.11, 0.15, 0.17 and 0.18 LCCO sample and $\Delta E$ ~ 68 meV for the $x$ = 0.13 and 0.175 LCCO and the $x$ = 0.15 NCCO sample. The RIXS polarization resolved measurements were conducted with a wider monochromator exit slit at a resolution of $\Delta E$ ~ 85 meV in order to partly compensate the reduced efficiency of the polarimeter with respect to the normal configuration. For each transferred momentum a non-resonant silver paint or carbon tape spectrum provided the exact position of the elastic (zero energy loss) line. For the polarimetric RIXS measurements of Figs. 1 (d) and (e), σ- polarization incident on the sample was used and the graded multilayer served as analyzer of the scattered photons as explained in details in Ref. 31.

In the theory calculations we consider the three-band Hubbard model with the following standard parameters in units of eV: $U_{dd}$ = 8.5, $U_{pp}$ = 4.1, $t_{pd}$ = 1.13, $t_{pp}$ =0.49, $\Delta_{pd}$ = 3.24 [32]. Determinant quantum Monte Carlo (DQMC) [33] is used to solve the model on a fully periodic 16 x 4 cluster at a temperature of $T$ = 0.125 eV. For each doping, 512 independently seeded Markov chains with 50000 measurements each are run. The charge susceptibilities obtained by DQMC are analytically continued to real frequency using the Maximum Entropy method (MEM) with model functions determined by the first moments of the data [34]. The inverse dielectric function plotted in Fig. 3(c) is obtained via $\frac{1}{\varepsilon(q,\omega)} = \frac{1}{1 + V_q \chi(q,\omega)}$, where $\chi(q,\omega)$ is the real frequency charge susceptibility from DQMC and MEM. The long-range and three dimensional Coulomb interactions neglected in the three-band Hubbard model are captured by $V_q$. We use the layered electron gas form [35]: $V_q = \frac{d}{2\varepsilon_\infty q} \frac{q \sinh(q\, d)}{\cosh(q\, d) - \cos(q_z d)}$, where $d$ is the interplane spacing and $q$ and $q_z$ are the in-plane and out-of-plane components of the momentum transfer, respectively. $\varepsilon_\infty$, the sole free parameter of our calculation, is adjusted to give a roughly 1eV mode for $q_z$ = 0 and the smallest nonzero $q$ = (0.0625, 0); its value is not varied with doping.

**Data availability**

The data that support the plots within this paper and other findings of this study are available from the corresponding authors upon reasonable request.

**Acknowledgments**

We thank J. Zaanen for helpful discussions. This work is supported by the U.S. Department of Energy (DOE), Office of Science, Basic Energy Sciences, Materials Sciences and Engineering Division, under contract DE-AC02-76SF00515. L.C. acknowledges the support from Department of Energy, SLAC Laboratory Directed Research and Development funder contract under DE-AC02-76SF00515. RIXS data were taken at the ID32 of the ESRF (Grenoble, France) using the ERIXS spectrometer designed jointly by the ESRF and Politecnico di Milano. G.G. and Y.Y.P. were supported by the by ERC-P-ReXS project (2016-0790) of the Fondazione CARIPLO and Regione Lombardia, in Italy. R.L.G and T.S. acknowledge support from NSF award DMR-1708334. Computational work was performed on the Sherlock cluster at Stanford University and on resources of the National Energy Research Scientific Computing Center, supported by the U.S. DOE under Contract No. DE-AC02-05CH11231.

## Author contributions

W.-S.L., G.G., L.B., T.P.D. and Z.-X.S. conceived the experiment. M.He., W.-S.L., L.C., R.F., Y.Y.P., G.G., M.Ha., K.K. and N.B.B. conducted the experiment at ESRF. M.He., L.C., and W.-S.L. analyzed the data. E.W.H., W.-C.L., B.M. and T.P.D. performed the theoretical calculations. T.S., J.H., C.R.R., Y.S.L. and R.L.G. synthesized and prepared samples for the experiments. M.He., B.M. and W.-S.L. wrote the manuscript with input from all authors.

## Additional information

## Competing financial interests

The authors declare no competing financial interests.

**Figure captions**

**Figure 1: Plasmons in a layered electron gas and dispersive charge excitations in electron-doped cuprates. a** Plasmon dispersion in a layered electron gas as a function of in- and out-of-plane momentum transfer $q_{//}$ and $q_z$, respectively. Different branches correspond to specific out-of-plane momentum transfers $q_z$ and are a result of the interplane Coulomb interaction between the periodically stacked planes with distance $d$ (see inset). Their spectrum varies from a single optical branch (dark red line) with the characteristic plasma frequency $\omega_p$ for $q_z = 0, 2\pi/d,...$ to a range of acoustic branches (light red lines) when $q_z = \pi/d, 3\pi/d,...$ is approached. The electron-hole pair excitation continuum is illustrated by the orange shaded area. **b, c** Resonant inelastic x-ray scattering (RIXS) spectra of NCCO ($x = 0.15$) and LCCO ($x = 0.175$) at in-plane momentum transfers $q_{//} = (0.045, 0)$ and $(0.095, 0)$ for incident photon energies tuned to the Cu $L_3$-edge at T ~ 20 K. The dispersive zone center excitation is highlighted by the red shaded peak profile. The additional peak at ~ 0.3 eV in **b** is the paramagnon. NCCO and LCCO spectra are offset in vertical direction for clarity. **d, e** Polarization resolved RIXS spectra of LCCO at $q_{//} = (0.045, 0)$ and $(0.095, 0)$. Charge excitations are detected in the parallel polarization channel ($\sigma\sigma$, red line) while magnetic excitations are detected in the spin flip crossed-polarization channel ($\sigma\pi$, blue line). The reference spectrum (open symbols) is taken with $\sigma$ polarized incident photons in absence of polarization analysis, corresponding thus to the sum of $\sigma\sigma$ and $\sigma\pi$. Red and blue markers indicate the relevant spectral weight maximum of the $\sigma\sigma$ and $\sigma\pi$ channel, respectively.

**Figure 2: Three dimensionality of the zone center excitations. a, b** RIXS intensity maps of LCCO ($x = 0.175$) for momentum transfer along the *hh*- and *h*- directions at $l = 1$ and $l = 1.65$. Red and gray symbols indicate least-square fit peak positions of the zone center excitation and the paramagnon, respectively (see Supplementary Figs. 5 and 6). Error bars are estimated from the uncertainty in energy loss reference point determination together with the error of the fits.

The insets indicate the probe direction in reciprocal space. **c** Summary of the energy dispersion of the zone center excitation (red symbols) and the paramagnon (gray symbols) for different $l$-values, *i.e.* different momentum transfers along the *c*-axis. The energy dispersion of the paramagnon is independent of $l$ within the experimental error. **d, e** Representative raw RIXS spectra (open symbols) for momentum transfer along the *hh*- and *h*- directions at different $l$-values, together with the anti-symmetrized Lorentzian fit profiles of the zone center excitation (red shades) and the sum of all contributions fitted to the spectra (solid black lines, see also Supplementary Figs. 5 and 6). Spectra are offset in vertical direction for clarity.

**Figure 3: Out-of-plane plasmon dispersion. a** RIXS intensity map of LCCO ($x = 0.175$) for momentum transfer along the out-of-plane direction at $h = 0.025$. The out-of-plane momentum is indicated by the indices $l$ (top scale, units of $2\pi/c$) corresponding to the crystallographic *c*-axis, and $l^*$ (bottom scale, units of $2\pi/d$) corresponding to the $CuO_2$ plane spacing. White symbols indicate fitted peak positions of the zone center excitation. **b** Crystal structure of LCCO with the crystallographic unit cell indicated by black lines and the $CuO_2$ planes in red. **c** Calculation of the charge dynamics for planes of a stacked three-band Hubbard model (see Methods) with electron doping $x = 0.18$. The peak frequency of the loss function (-Im $1/\varepsilon$) is plotted vs. out-of-plane momentum transfer $l^*$ at $h = 0.1875, 0.1250,$ and $0.0625$. The energy is expressed in units of $t_{pd}$, the hopping integral between the oxygen *2p* and Cu *3d* orbitals in the three-band Hubbard model.

**Figure 4: Doping dependence of the plasmon. a** Energy dispersion of the LCCO zone center mode for momentum transfer along the *h*-direction at $l^* = 0.5$. Red symbols are the fitted peak positions for Ce doping concentrations $x = 0.11, 0.13, 0.15, 0.17,$ and $0.18$. The bottom inset shows the temperature vs. doping phase diagram of LCCO thin films adapted from Ref. 29

including a superconducting dome (orange shade, SC) and a small Fermi surface region (blue shade) undergoing a crossover to a reconstructed Fermi surface (blue line, FSR) that ends at $x \sim 0.14$. **b** Representative raw RIXS spectra (open symbols) for momentum transfer $h = 0.035$ and $l^* = 0.5$ for different dopings. The anti-symmetrized Lorentzian fit profiles are shaded in red with markers indicating the peak positions. Spectra are offset in vertical direction for clarity. **c** Mode energies vs. the square root of $x$ for in-plane momenta from $h = 0.035$ to $h = 0.095$ at $l^* = 0.5$. The symbol style refers to the dopings as indicated in panel **a.** Dashed lines are linear fits of the mode energy vs. $\sqrt{x}$ for dopings $x = 0.11$ to $0.15$, assuming the mode energy is 0 at $x = 0$.

# Figures

# Figure 1

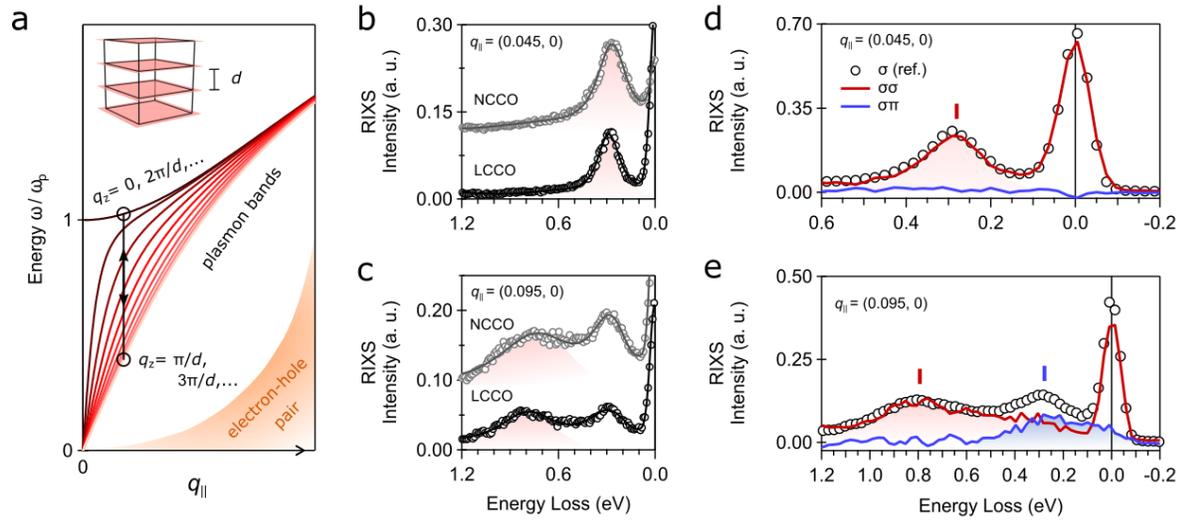

# Figure 2

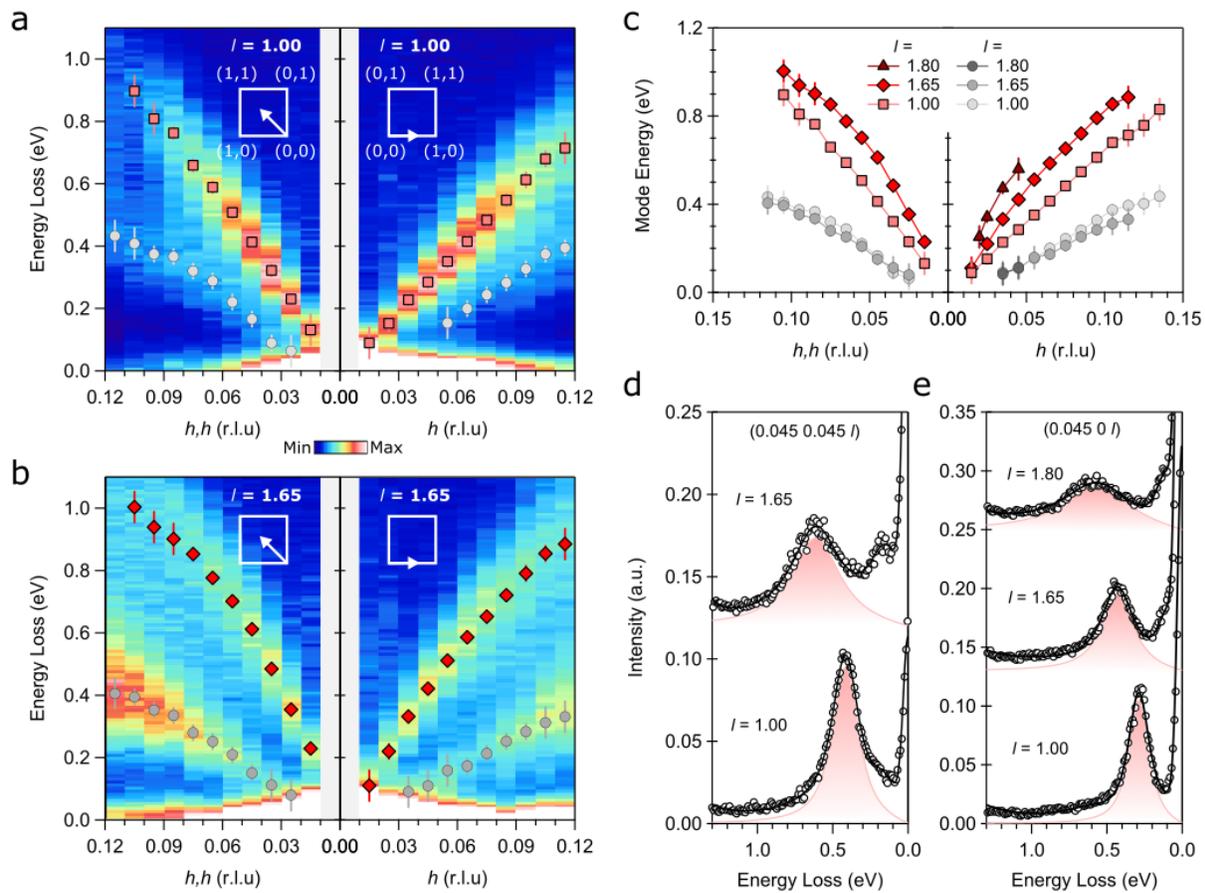

Figure 3

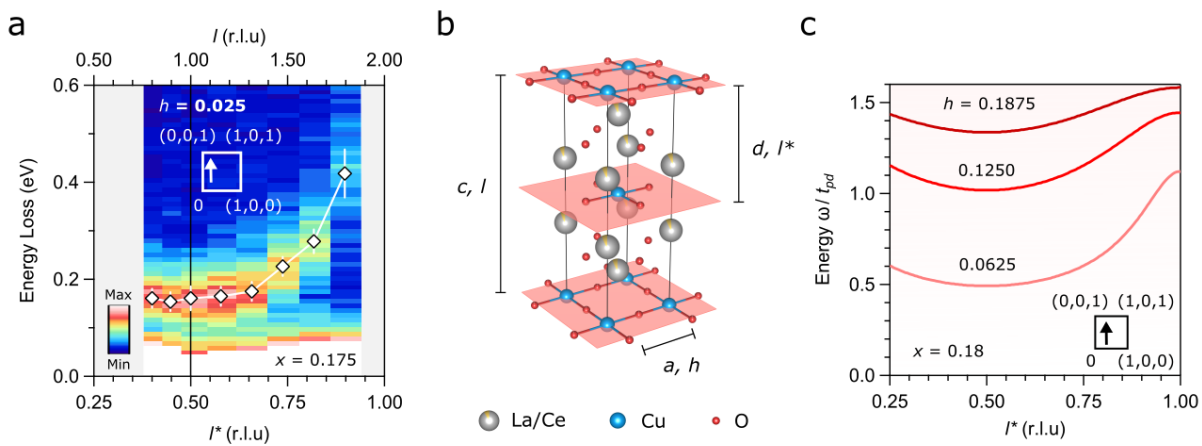

Figure 4

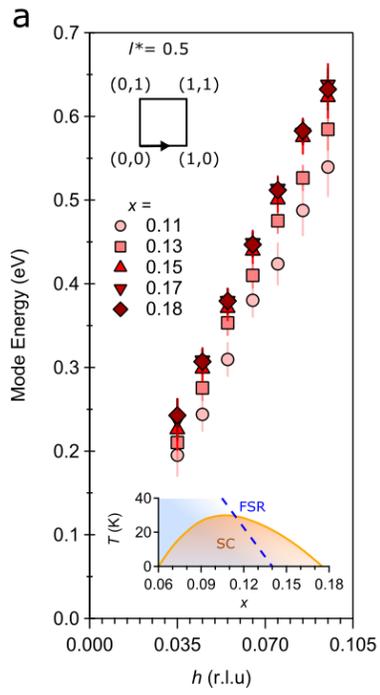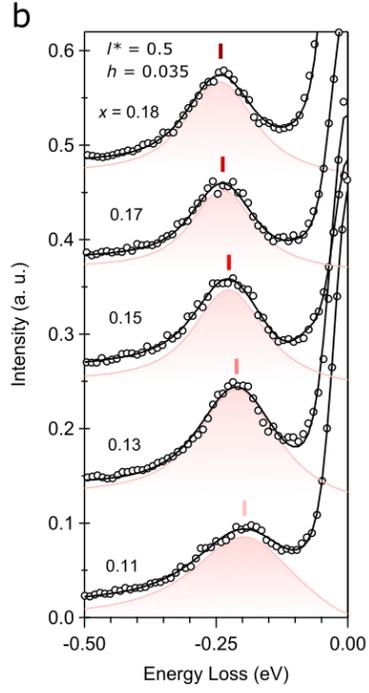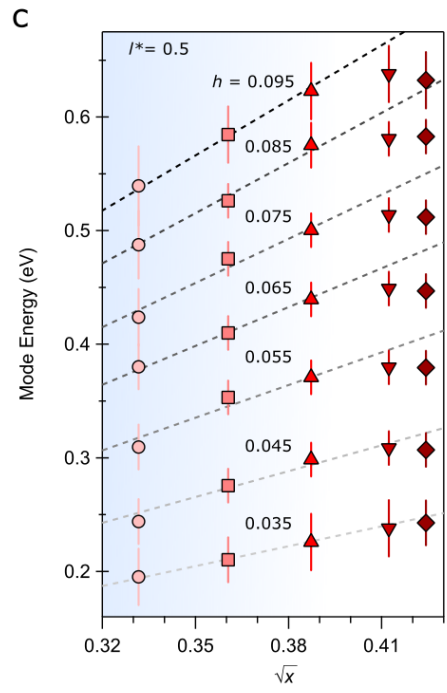

# Three dimensional collective charge excitations in electron-doped cuprate superconductors

M. Hepting, L. Chaix, E. W. Huang, R. Fumagalli, Y. Y. Peng, B. Moritz, K. Kummer, N. B. Brookes, W. C. Lee, M. Hashimoto, T. Sarkar, J. F. He, C. R. Rotundu, Y. S. Lee, R. L. Greene, L. Braicovich, G. Ghiringhelli, Z. X. Shen[*], T. P. Devereaux[*], and W. S. Lee[*]

Correspondence to: leews@stanford.edu, zxshen@stanford.edu, tpd@stanford.edu

10/28/18

**This PDF file includes:**

Supplementary Text
Supplementary Figures
Supplementary References

**Supplementary Text**

**Fit of the plasmon dispersion in the layered electron gas model**

We consider the energy-momentum dispersion of a plasmon mode in a layered electron gas model for momentum transfer $q$ along the $h$-direction at fixed out-of-plane momentum transfer values $q_z$ (along the $l^*$-direction). Let $d$ be the spacing between adjacent $CuO_2$ planes and $q_z\, d = \pi\, l^*$ and $\varepsilon_\infty$ the high-frequency dielectric constant due to the screening by the core electrons. Following Supplementary Ref. 1, the Coulomb potential of a layered electron gas is

$$V_q = \alpha^2\, \frac{\sinh(q\, d)}{q\, [\cosh(q\, d) - \cos(q_z\, d)]}$$

with

$$\alpha = \sqrt{\frac{e^2 d}{2\, \varepsilon_0 \varepsilon_\infty q}}.$$

In an isotropic medium it is well known that the three-dimensional Coulomb potential is $e^2/\varepsilon_0\varepsilon_\infty q^2$, while in a two-dimensional plane the Coulomb potential is $e^2/2\varepsilon_0\varepsilon_\infty q$. These are the two limits of the above form of the layered electron gas with $V_q$ becoming $V_q = e^2/\varepsilon_0\varepsilon_\infty q^2$ in the approximation of long-wavelengths ($q_z\, d \ll 1$ and $q\, d \ll 1$), and $V_q = e^2/2\varepsilon_0\varepsilon_\infty q$ for short wavelengths ($q\, d \gg 1$, independent of $q_z$ momentum).

Here, we use the full form from above and obtain the energy of the plasmon mode by

$$\omega_p \sim q\sqrt{V_q} \sim \alpha \sqrt{\frac{q\, \sinh(q\, d)}{\cosh(q\, d) - \cos(q_z d)}}$$

With this expression we performed least square fits of the plasmon energy dispersion, simultaneously for the data sets of $l^* = 0.5$, $l^* = 0.825$ and $l^* = 0.9$ with the global fitting parameter $\alpha$, yielding $\alpha = 2.06$ [Supplementary Fig. 8]. We note that the layered electron gas model does not include strong correlations which can explain deviations between the experimental data and the model fit.

**Supplementary Figures**

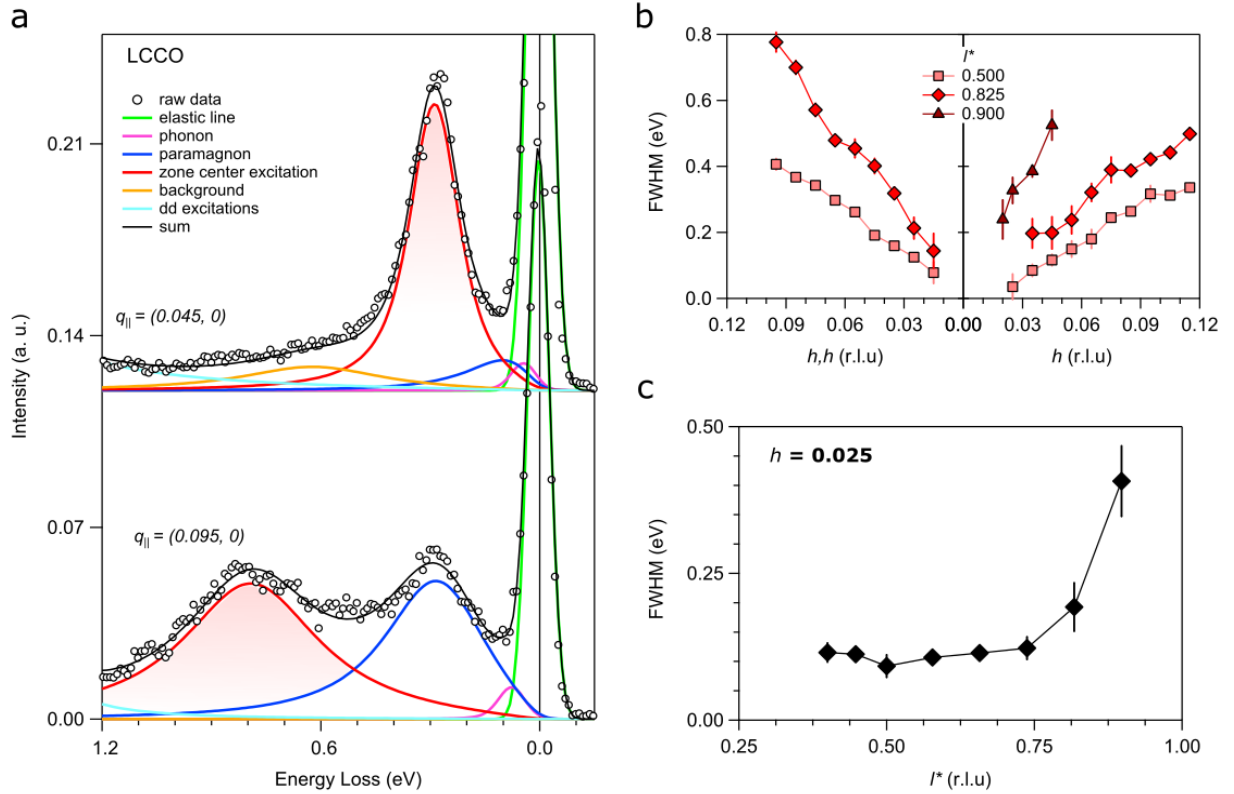

**Supplementary Figure 5: Fits of the RIXS spectra. a** Fits of LCCO ($x = 0.175$) RIXS spectra at in-plane momentum transfer positions q// = (0.045 0) and (0.095 0), representative for all fits performed in the scope of this work. The model uses a Gaussian for the elastic peak (green) and anti-symmetrized Lorentzians for all other contributions in the spectrum, convoluted with the energy resolution (here $\Delta E = 68$ meV) via Gaussian convolution. The peak profiles of the zone center excitation (plasmon) are shaded in red. **b** FWHM of the zone center excitation (plasmon) as extracted from the fits for momentum transfer along the *hh*- and *h*- directions at $l^* = 0.5$, $l^* = 0.825$ and $l^* = 0.9$, corresponding to the fitted peak positions shown in Fig. 2(c) of the main text. Error bars are from the fits. **c** FWHM of the zone center excitation (plasmon) as extracted from the fits for momentum transfer along the out-of-plane direction at $h = 0.025$. The panel corresponds to the fitted peak positions shown in Fig. 3(a) of the main text.

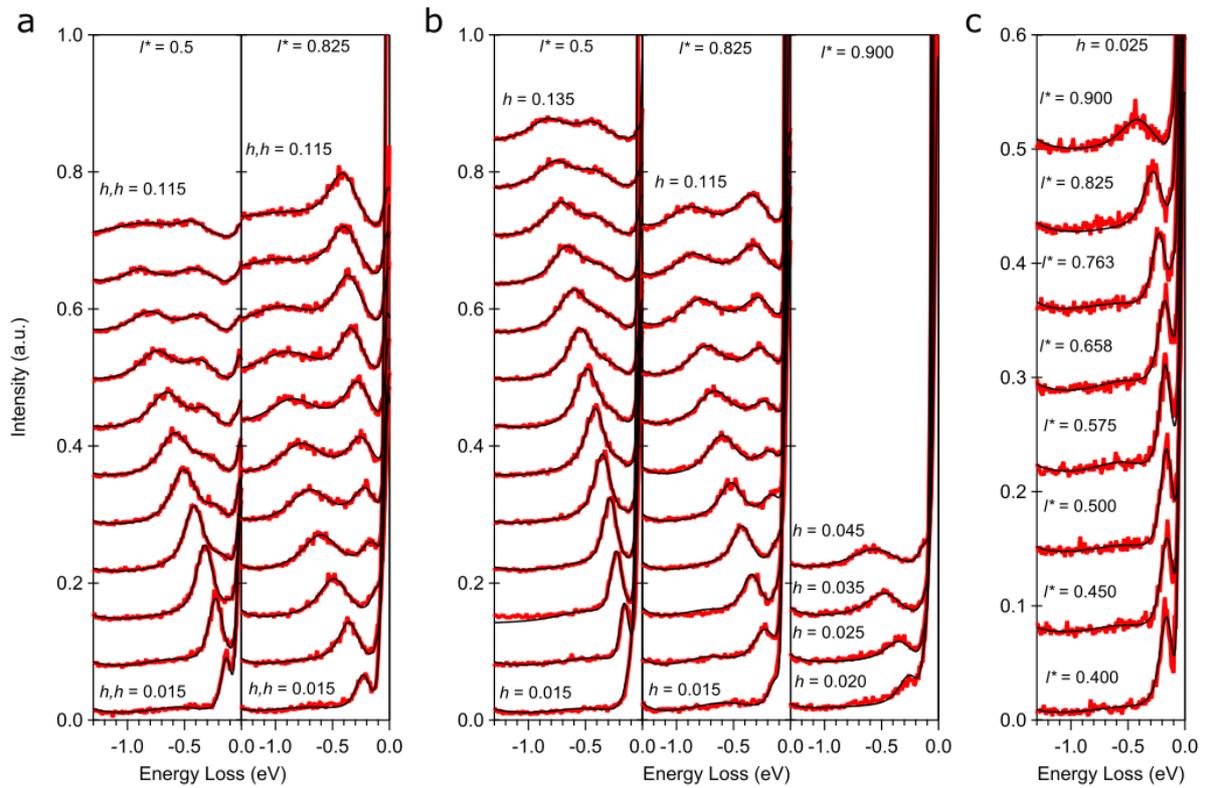

**Supplementary Figure 6: Raw data and fits of the RIXS spectra. a, b** Raw RIXS spectra (red) of LCCO ($x = 0.175$) together with the fits (solid black lines) for momentum transfer along the *hh*-direction (**a**) and *h*-direction (**b**) at different *l\**. The spectra are offset in vertical direction for clarity. **c** Raw RIXS spectra together with the fits for momentum transfer along the *l\**-direction at $h = 0.025$.

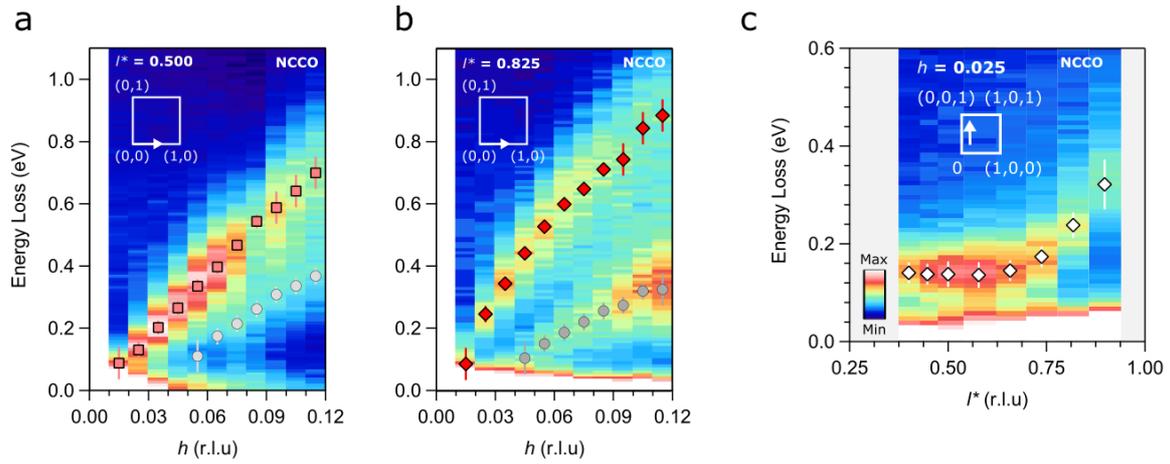

**Supplementary Figure 7: Three dimensionality of the zone center excitations in NCCO.**
**a,b** RIXS intensity maps of NCCO ($x$=0.15) for momentum transfer along the $h$-direction at $l^* = 0.5$ and $l^* = 0.825$. Red and gray symbols indicate least-square fit peak positions of the zone center excitation and the paramagnon, respectively. The inset indicates the probe direction in reciprocal space. **c** RIXS intensity map of NCCO ($x = 0.15$) for momentum transfer along the out-of-plane direction at $h = 0.025$. White symbols indicate fitted peak positions of the zone center excitation.

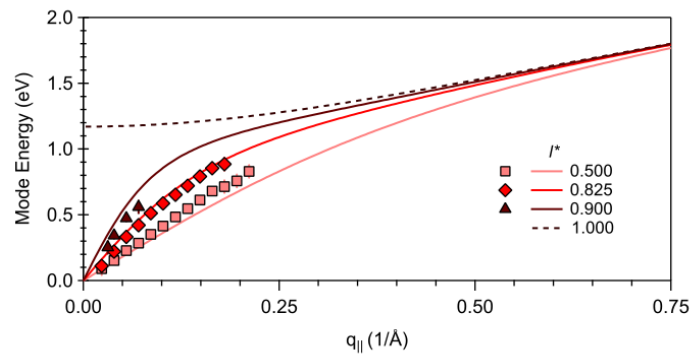

**Supplementary Figure 8: Fits of the plasmon dispersion in the layered electron gas model. a** Fits (solid lines) of the mode energies of LCCO ($x = 0.175$) (red symbols) as function of in-plane momentum transfer $q_{//}$ along the $h$- direction at $l^* = 0.5$, $l^* = 0.825$ and $l^*$

= 0.9. The fit is global, *i.e.* the three *l\** data sets are fitted simultaneously with the same fit parameter, as described in the Supplementary text.

**Supplementary References**